\documentclass[useAMS,usenatbib]{mn2e}
\usepackage{times,graphicx,hyperref}
\usepackage{amssymb}

\def\farcs{\hbox{$.\!\!^{\prime\prime}$}}

\title[BSs in Sagittarius globular clusters]{Structural parameters and blue stragglers in Sagittarius dwarf spheroidal galaxy globular clusters\thanks{Based
    on data collected at the Danish telescope at the ESO-La Silla
    observatory.}}

\author[R. Salinas et al.]
{Ricardo Salinas$^{1}$\thanks{E-mail: rcsave@utu.fi}, 
Lucie J\'ilkov\'a$^{2,3}$, 
Giovanni Carraro$^{2,4}$,   
M\'arcio Catelan$^{5}$, and
P\'{i}a Amigo$^{5}$ \\
$^{1}$Finnish Centre for Astronomy with ESO (FINCA), University of Turku, V\"ais\"al\"antie 20, FI-21500 Piikki\"o, Finland\\
$^{2}$European Southern Observatory, 
Alonso de C\'ordova 3107, Casilla 19001, Santiago 19, Chile\\
$^{3}$Department of Theoretical Physics and Astrophysics, 
Faculty of Science, Masaryk University, 
Kotl\'a\v rsk\'a 2, CZ-611 37 Brno, Czech Republic\\
$^{4}$Dipartimento di Astronomia, Universit\'a di Padova,
Vicolo Osservatorio 5, I-35122, Padova, Italy\\
$^{5}$Departamento de Astronom\'ia y Astrof\'isica, Pontificia
Universidad Cat\'olica de Chile, Av. Vicu\~na Mackenna 4860, 782-0436 
Macul, Santiago, Chile}

\begin{document}

\date{Accepted 2011 December 7.  Received 2011 November 15; in original form 2011 October 12.}

\pagerange{\pageref{firstpage}--\pageref{lastpage}} \pubyear{2011}

\maketitle

\label{firstpage}
%%%%%%%%%%%%%%%%%%%%%%%%%%%%%%%%%%%%%%%%%%%%%%%%
\begin{abstract}
  We present $BV$ photometry of four Sagittarius dwarf spheroidal
  galaxy globular clusters: Arp~2, NGC~5634, Palomar~12, and Terzan~8,
  obtained with the Danish telescope at ESO-La Silla.  We measure the
  structural parameters of the clusters using a King profile fitting,
  obtaining the first reliable measurements of the tidal radius of
  Arp~2 and Terzan~8. These two clusters are remarkably extended and
  with low concentrations; with a concentration of only $c=0.41 \pm
  0.02$, Terzan~8 is less concentrated than any cluster in our Galaxy.
  
  Blue stragglers are identified in the four clusters, and their spatial 
  distribution is compared to those of horizontal branch and red giant branch 
  stars. The blue straggler properties do not provide evidence of mass
  segregation in Terzan~8, while Arp~2 probably shares the same
  status, although with less confidence. In the case of NGC~5634 and
  Palomar~12, blue stragglers are significantly less populous, and their
  analysis suggests that the two clusters have probably undergone mass 
  segregation.
\end{abstract}


%%%%%%%%%%%%%%%%%%%%%%%%%%%%%%%%%%%%%%%%%%%%%%%%
\section{Introduction}
\label{sec:intro}

Stars known as blue stragglers (BSs) occupy a position on the
color-magnitude diagram (CMD) of stellar clusters at a higher
luminosity and temperature than the main-sequence (MS) turnoff point,
appearing as a sparse prolongation of the MS
\citep[e.g.,][]{sandage53,burbidge58}. This position implies that
these stars have a larger mass than the one expected from their parent
cluster evolution. Two different, but non-exclusive
\citep[e.g.,][]{ferraro09}, scenarios have been postulated for BS
formation, namely mass exchange or the merger of close binaries
\citep[e.g.,][]{mccrea64,carney01} and the direct collision of stars
\citep[e.g.,][]{hills76,leonard89,davies94,leonard95}, which is
naturally enhanced in crowded environments.

The formation of BSs is then intimately related to the structure and
dynamical status of their surroundings. In this respect the
Sagittarius dwarf spheroidal galaxy (hereafter Sgr), due to its
ongoing disruption into our Galaxy, offers different environments
where the BS phenomenon can be studied.  Since the discovery of Sgr
\citep{ibata94}, a plethora of globular clusters have been putatively
associated with it. Four of them, M54 (NGC~6715), Arp~2, Terzan 7 and
Terzan~8, lying within the body of Sgr, are typically considered as
bona fide members \citep{dacosta95}. Using several arguments,
including dynamical modeling, the position over the stream of stars
from the disrupting Sgr and abundance ratios, several other globular
clusters have been considered as (former) members of the Sgr cluster
system that have been stripped away due to tidal interactions with the
Galaxy
\citep{dinescu00,palma02,bellazzini03,cohen04,carraro07,law10,forbes10}.

In the present paper we study the BS population in four of
the Sgr globular clusters: the bona fide members Arp~2 and Terzan~8,
and the stripped NGC~5634 and Palomar~12. These clusters cover a range of ages,
metallicities, galactocentric distances and structural properties, which 
however have heretofore been scarcely studied with appropriate data.

Arp~2 (${\rm [Fe/H]}=-1.83$; \citealt{mottini08}) is a particularly
remarkable cluster, since despite several attempts, there is still no
agreement whether it has a somewhat younger age
\citep{buonanno95,marin09} or the same as the bulk of the Galactic
globular clusters \citep{layden00,dotter10}. A young age would clash with its
mostly blue horizontal branch \citep[HB;][]{buonanno95}. Terzan~8, also lying
in the Sgr body, is the most metal-poor (${\rm [Fe/H]}=-2.34$;
\citealt{mottini08}), and also perhaps the oldest, cluster in Sgr
\citep{montegriffo98,dotter10}. NGC~5634 is much like Terzan~8 in
terms of age and metallicity \citep{bellazzini02} and has been
associated to Sgr given its position and radial velocity
\citep{bellazzini02,law10}. The final cluster in this study, 
Palomar~12, has been associated with Sgr due to its proper motion
\citep{dinescu00} and anomalous abundances \citep{cohen04}. In addition, 
traces of the Sgr stream have been found in its surroundings
\citep{martinez02}. Palomar~12 is a relatively young \citep[$\sim$\,8--9\,Gyr;][]{stetson89,rosenberg98} and metal-rich (${\rm [Fe/H]}=-1.0$,
\citealt{brown97}) cluster.

The paper is organized as follows: we present our observations and
reduction procedures in Sect.~\ref{sec:obs}; the measurement of the
structural parameters of the clusters is given in
Sect.~\ref{sec:structure}. We present the results on the BS spatial 
distribution in Sect.~\ref{sec:stragglers}. Finally, our main results 
concerning the BS spatial distribution and dynamical status of the 
clusters in our sample are discussed in Sect.~\ref{sec:discussion}, 
and a brief summary is presented in Sect.~\ref{sec:summary}.

%%%%%%%%%%%%%%%%%%%%%%%%%%%%%%%%%%%%%%%%%%%%%%%%
\section {Observations and data reduction}
\label{sec:obs}

\begin{figure}
\includegraphics[width=0.49\textwidth]{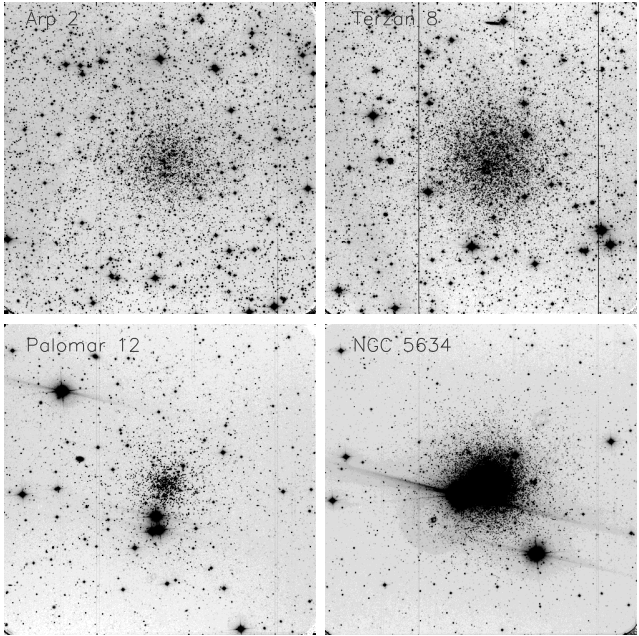}
\caption{Medianed images produced with MONTAGE2. These images are used
  to establish the initial positions of the stars that are introduced
  to ALLFRAME. All images have approximately
  $14\arcmin\times14\arcmin$. North is up and East to the left.}
\label{fig:clusters}
\end{figure}

\begin{figure}
\includegraphics[width=0.49\textwidth]{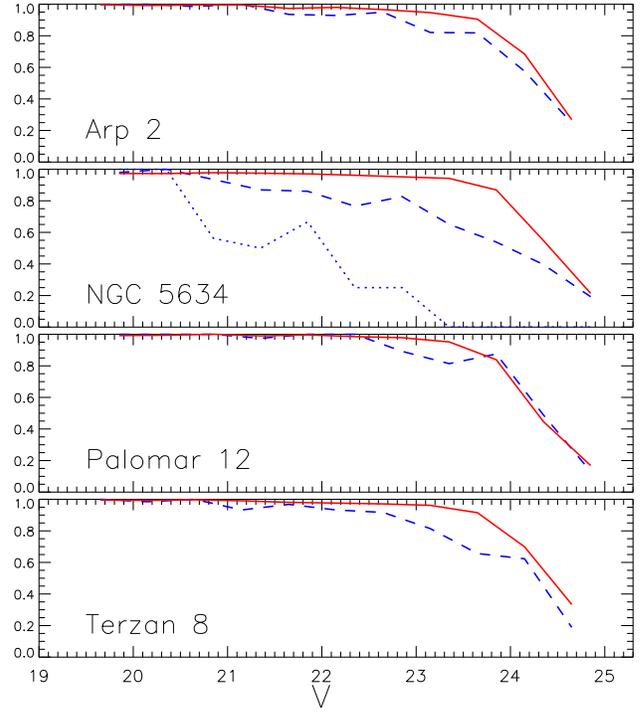}
\caption{Level of completeness for the $V$ photometry in each
  cluster. Stars were counted in 0.5\,magnitude bins. The dashed blue
  line indicates the completeness inside the inner 2\arcmin of each
  cluster, while the red solid line indicates the completeness outside
  this radius. NGC~5634 is divided into three bins with
  $30\arcsec<R<60\arcsec$ (dotted line), $60\arcsec<R<120\arcsec$
  (dashed line), and $R>120\arcsec$ (solid line).}
  \label{fig:completeness}
\end{figure}

Bessel $B$ and $V$ images were obtained using the Danish Faint Object
Spectrograph and Camera (DFOSC) mounted on the 1.54\,m Danish Telescope
at ESO La Silla between June 27 and 30, 2003. A series of short
exposures were obtained ranging from 200 to 600 seconds in the $B$
band and 100 to 300\,seconds using the $V$ filter. The seeing quality
was good during the four nights, with FWHM in the range
0.8\arcsec--1.3\arcsec. The DFOSC field of view is
13.5\arcmin$\times$13.5\arcmin\, with a scale of $0\farcs39$
pixel$^{-1}$.

\begin{table*}
 \centering
  \caption{Exposure times and seeing conditions for the images used.}
\label{table:seeing}  
\begin{tabular}{@{}lllc@{}}
  \hline
Cluster name & $B$ exp. time &      $V$ exp. time &  Seeing range\\
&[seconds]&[seconds]&[$\arcsec$]\\
\hline
NGC~5634& 6$\times$250+1$\times$300+4$\times$400+&6$\times$100+1$\times$120+1$\times$150+ & $B$: 0.9--1.1\\
 & 1$\times$450+2$\times$500+1$\times$600&4$\times$200+2$\times$250+1$\times$300& $V$: 0.8--0.9\\
Arp~2 & 6$\times$250+9$\times$400&5$\times$120+1$\times$100+9$\times$200& $B$: 0.9--1.0\\
& & &$V$: 0.8--0.9\\
Terzan~8 & 5$\times$200+4$\times$250+6$\times$400&5$\times$80+4$\times$120+6$\times$200&$B$: 0.9--1.0\\
& & &$V$: 0.8--1.0\\
Palomar~12& 2$\times$250+1$\times$300+7$\times$400+&1$\times$120+1$\times$150+9$\times$200+&$B$: 0.9--1.0\\
& 1$\times$450+4$\times$500& 2$\times$220+2$\times$250&$V$: 0.9--1.0\\
\hline
\end{tabular}
\end{table*}

Reduction of the images, including bias and overscan subtraction and
flat-fielding, was done using standard tasks within
IRAF.\footnote{IRAF is distributed by the National Optical Astronomy
  Observatory, which is operated by the Association of Universities
  for Research in Astronomy (AURA) under cooperative agreement with
  the National Science Foundation.}  Point-spread function (PSF)
photometry for all the images was done using DAOPHOT/ALLSTAR
\citep{stetson87}, where typically $\sim$50 bright and isolated stars
were selected in the NGC~5634 and Pal~12 fields and $\sim$100 stars in
the more populated Arp~2 and Terzan~8 fields (see
Figure~\ref{fig:clusters}) in order to derive the PSF.  For the cluster
photometry, the best 15 images in each band for each cluster were
selected. Details of exposure time and image quality can be seen in
Table \ref{table:seeing}. The entire set of images has been used for a
reassessment of the clusters' stellar variability, as will be
presented in a forthcoming paper, extending the results from
\citet{salinas05}.

Transformation equations in pixel space for the positions of the stars
between the images were derived using DAOMATCH/DAOMASTER
\citep{stetson93}. This step requires the selection of a ``master
image'' to which transformation equations are anchored to. The master
image for each cluster was selected to be the one with the best seeing
and longer exposure. The subset of images for each cluster from both
filters was combined together using MONTAGE2 \citep{stetson94}, which
weighs each image according to its seeing and total flux (after sky 
subtraction), hence making sharper and deeper images more
preponderant in the final combined frame. From these combined images,
a master list of stars was generated by applying iteratively
DAOPHOT/ALLSTAR. This list of stars together with the images were then
fed into ALLFRAME \citep{stetson94}, which fits simultaneously the
PSF to all stars in all the available images. Mean
magnitudes and errors for each band from the output ALLFRAME photometry
were derived again using DAOMATCH/DAOMASTER, which match the star
coordinates in the different frames and apply a zero-point correction
to the magnitudes respect to the master image.

\begin{figure*}
\includegraphics[width=\textwidth]{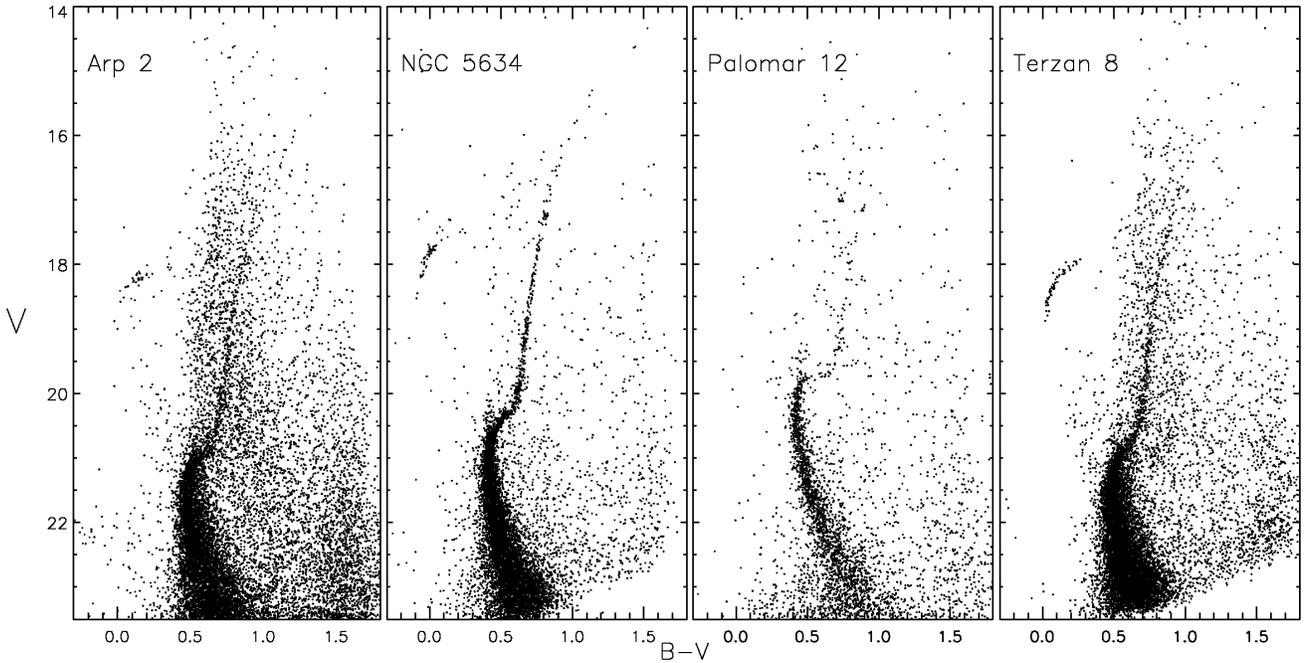}
\caption{Color-magnitude diagrams of the four clusters. Field
  contamination can be seen as most severe in the case of Arp~2.}
\label{fig:cmds}
\end{figure*}

\citet{landolt92} standard stars were observed in the fields Mark~A,
G26 and PG~1525 during the first, second and fourth nights of
observations. These fields were observed at a range of airmasses,
$1\!<\!X\!<\!1.4$, bracketing the science exposures. Transformation
equations to the standard system were derived using
IRAF/PHOTCAL. Aperture corrections, also calculated within the
IRAF/PHOTCAL environment, were found to be always below $-0.02$\,mag for
both filters in all clusters.

Globular clusters studies using ALLFRAME photometry usually deal with
large image sets comprising hundreds if not thousands of images
\citep[e.g.,][]{walker11}. Artificial stars experiments hence become
prohibitive and completeness has to be estimated using the behaviour
of the faint end of the observed luminosity function. In our case, the
modest amount of images allows us to study the completeness directly
with artificial stars experiments. In our procedure we included 1000
artificial stars in random positions for each cluster, except for
Palomar~12 where 600 stars were added. The magnitudes of these
artificial stars were between $19.5<V<25$, i.e. roughly from the BS 
magnitude level to the level of the faintest stars detected with
ALLFRAME. These artificial stars were initially included in each
cluster master image and then into the rest of the image set by using
the inverse of the transformation equations derived with DAOMASTER. In
this way, artificial stars are placed in the exact same relative
position inside each frame. For each cluster 10 new image sets were
generated in this way, each having 15 $B$ and 15 $V$ images like the
original set. Each new image set was photometered with ALLFRAME in the
same way as the original image set, i.e. the one with no additional
stars added.

The level of completeness for each cluster can be seen in
Figure~\ref{fig:completeness}. In Arp~2, Palomar~12 and Terzan~8, which
have uncrowded fields, we analyzed the completeness in two
areas. Beyond 2\arcmin\, the photometry of these three clusters can be
considered as 100\% complete to around $V\!\sim\!23$ (solid line), while
inside this radius, 100\% is reached down to $V\!\sim\!22.5$. For 
NGC~5634, which provides us with the most crowded central fields (see
Figure~\ref{fig:clusters}), we made the count of artificial stars in
three concentric annuli, with $30\arcsec<R<60\arcsec$,
$60\arcsec<R<120\arcsec$, and $R>120\arcsec$; the level of
completeness in these areas is depicted in Figure~\ref{fig:completeness}
as the dotted, dashed and solid lines, respectively. The incompleteness 
in the inner parts is especially severe, with 100\% being reached only 
down to $V\sim20.5$) but in any case not affecting the BS magnitude 
level. 

In conclusion, the photometry can be considered as complete in the
entire magnitude range where BSs are present in the four clusters,
with the exception of the inner 30\arcsec of NGC~5634, where at
$V\sim19.5$ the completeness level is only about 50\%. This central
area is then excluded from our analysis. Final CMDs of the four clusters can be seen in
Fig.~\ref{fig:cmds}.

\begin{figure*}
\centering
\includegraphics[width=0.9\textwidth,angle=270]{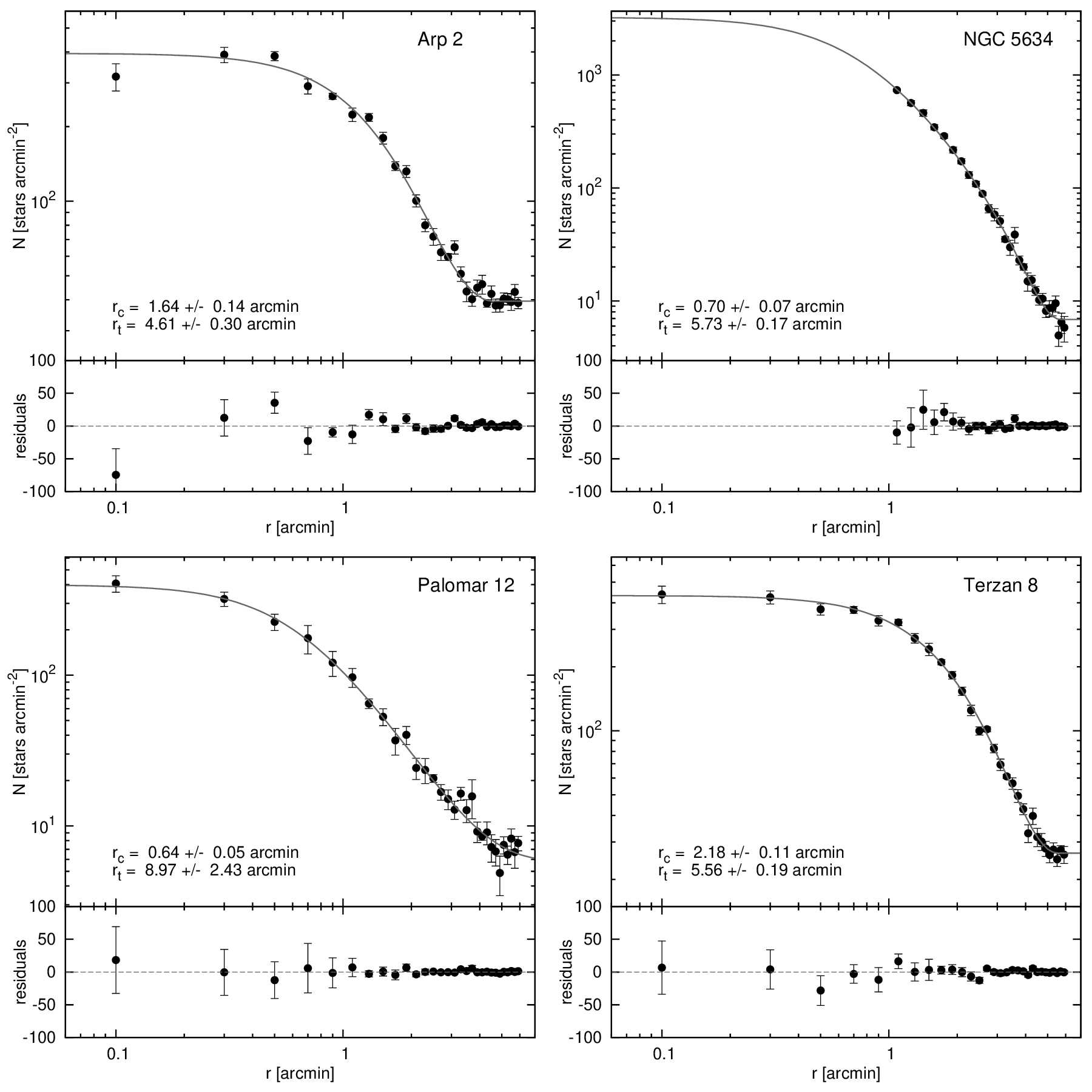}
\caption{Surface stellar density profile for each of our four
  clusters. The solid lines show the best-fitting King profiles, while
  the lower panels show the fit residuals.}
\label{fig:structure}
\end{figure*}

%%%%%%%%%%%%%%%%%%%%%%%%%%%%%%%%%%%%%%%%%%%%%%%%
\section{Structural parameters}
\label{sec:structure}

Structural parameters of the Sgr dSph globular clusters have been
measured over small-sized detectors and inhomogeneous samples
\citep[e.g.,][]{webbink85,trager95,mclaughlin05} or using HST images
\citep{mackey03c}, which are equally unsuited for extended clusters as
Arp~2 and Terzan~8. Taking advantage of the larger field of view
presented in this work, we re-derive the structure parameters of the
four clusters, using a \citet{king62} profile fitting. The estimation
of the dynamical time scales in a globular cluster depends on an
reliable measurement of the cluster core radius. The link between
cluster dynamical quantities and the BS population is explored in the
discussion in Sect.~\ref{sec:discussion}.

As a first step, the astrometry of each cluster is obtained by
cross-correlation with bright point sources in 2MASS catalogs. For the
analysis of stellar surface density we apply a completeness
magnitude limit of $V\!=\!23$ for all clusters. We calculate the
center of the clusters iteratively by averaging stellar
coordinates. In the first step, the average coordinates of stars
within a ``tidal'' radius $r_{\mathrm{max}}$ of 6\arcmin\,--\,the
maximum radius giving an aperture within the field of view\,--\,around
a trial center \citep[coordinates from][]{harris96} give a new cluster
center and its uncertainties. This new center is used iteratively with
the $r_{\mathrm{max}}$ trial ``tidal'' radius, until the central
coordinates stop changing, within their uncertainties.

\begin{table*}
 \centering
 \caption{Structural parameters from the King profile fitting and derived quantities. For the absolute magnitudes we have used distances and reddening from the 2010 edition of \citet{harris96}.}
\label{table:king}  
\begin{tabular}{@{}llllllllllll@{}}
  \hline
  Cluster &$\alpha$ center& $\Delta \alpha$&$\delta$ center&$\Delta \delta$&$D_{\rm Sgr}$ &$r_{\rm c}$&$r_{\rm t}$&$c$ &$r_{\rm h}$ &$M_V$&$t_{\mathrm{rc}}$\\
  &[deg, J2000]&[\arcsec]&[deg]&[\arcsec]&[deg]&[arcmin]&[arcmin]&&[arcmin]&&[Gyr]\\
  \hline
  Arp~2     & 292.181381 & 1.8  &   $-30.356551$ & 1.5 & 8.42  &  $1.64\pm0.14$ &   $4.61\pm0.30$&$0.45\pm0.05$ &1.65 &$-5.41$ & 7.9  \\
  NGC~5634  & 217.404460 & 0.9  &   $-5.9773490$ & 0.9 & 70.74 &  $0.70\pm0.07$ &   $5.73\pm0.17$&$0.91\pm0.05$ &0.85 &$-7.23$ & 11.9 \\
  Palomar~12& 326.665521 & 3.6  &   $-21.248884$ & 3.3 & 43.88 &  $0.64\pm0.05$ &   $8.97\pm2.43$&$1.15\pm0.12$ &0.97 &$-4.15$ & 0.56 \\
  Terzan~8  & 295.432784 & 1.7  &   $-33.999284$ & 1.4 & 12.19 &  $2.18\pm0.10$ &   $5.56\pm0.19$&$0.41\pm0.02$ &2.00 &$-5.68$ & 16.6 \\
  \hline
\end{tabular}
\end{table*}

The radial profile of the stellar surface density is obtained following
\citet{djorgovski88}. The cluster within the radius $r_{\mathrm{max}}$
is divided into 30 annular apertures with equal spacing. Each aperture
is subdivided into eight sectors defined by wedges of 45$^\circ$ angles
centered on the cluster. Star counts are made in each sector and then
averaged to obtain a mean stellar surface density and its uncertainty
at the middle radius of each annulus.

Having the radial profile of surface stellar density we fit the \citet{king62}
profile,  

\begin{equation}
n(r) = k\left\{ \frac{1}{\sqrt{1+(r/r_{\mathrm{c}})^2}} -
    \frac{1}{\sqrt{1+(r_{\mathrm{t}}/r_{\mathrm{c}})^2}}\right\}^2 + b
\end{equation}

\noindent for $r<r_{\mathrm{t}}$, and $n(r)=b$ for $r \ge r_{\mathrm{t}}$~-- where
$r_{\mathrm{c}}$ is the core radius, $r_{\mathrm{t}}$ the tidal
radius, $k$ a scaling constant, and $b$ the background level, fitted
via a non-linear least-squares method. A new cluster center is then 
found using the new value of $r_{\mathrm{t}}$. The new center gives a next
iteration of radial profile with new $r_{\mathrm{t}}$. We iterate
until the values of cluster's center and $r_{\mathrm{t}}$ stop 
changing, within their uncertainties. The convergence is basically
achieved within two iterations for all four clusters.

The construction of the radial profile of NGC~5634 was treated in a
special way. A strong contamination due to a bright star close to the
cluster center (Figure~\ref{fig:clusters})
is present in the star counts. To avoid this contamination, 
we created the stellar surface density profile using only stars with 
$r>1\arcmin$. 
 
The final stellar number density profiles with the best-fitting King
profiles are shown in Figure~\ref{fig:structure}. Structural
parameters are given in Table~\ref{table:king}. The derived
coordinates of the center of the clusters are in good agreement with
the ones measured by \citet{goldsbury10} using HST data and with the
\citet{harris96} catalogue value for NGC~5634 (not included in
\citeauthor{goldsbury10}). We also provide new estimations of the
absolute magnitude of the clusters. These are obtained by summing the
light contribution from all the stars within the tidal radius that are
consistent with being cluster members based on their position on the
CMDs.

The clusters still belonging to Sgr, Arp~2 and Terzan~8, show clear
differences with the ones that have been lost into the Galactic halo,
NGC~5634 and Palomar~12. While the former have large core radii and
very low concentrations, the latter show the opposite trend. Tidal
radii for all clusters but Palomar~12 are inside the radius of the
maximum allowed aperture ($r_{\mathrm{max}}$ above), so no
extrapolation of the data is necessary to determine $r_{\mathrm{t}}$.

The clusters lying inside Sgr deserve special remarks. With a
concentration of $c=0.45$ and $r_{\mathrm{c}}=13.6$\,pc, Arp~2 is
among the clusters with lowest concentration, when compared to the
Galactic globular clusters as given in \citet{harris96}. The existence
of two RR Lyrae stars beyond its tidal radius \citep{salinas05} may be
an indication of tidal disruption, although contamination of RRL stars
from Sgr is also a likely explanation (see below).

The parameters of Terzan~8 are even more remarkable. The cluster has a core
radius of 131\arcsec (16.5\,pc at the cluster distance), which is not only
the largest among the Sgr globulars, but also among the clusters in the 
Fornax dSph and the Magellanic Clouds \citep{mackey03a,mackey03c}. A
comparison with Galactic globular clusters core radii in the
\citet{harris96} catalogue reveals that only Palomar~14 possesses a
larger core radius, with $r_{\mathrm{c}}=18.2$\,pc~-- although new
measurements by \citet{sollima11} imply a core radius of about
12.4\,pc. This large core radius makes Terzan~8 join the select
group of clusters in our Galaxy with $r_{\mathrm{c}}>15$\,pc, which 
also includes 
Palomar~5, Palomar~14, and Palomar~15. The first two have revealed
clear evidence of tidal disruption \citep{odenkirchen01,sollima11},
while this is also suspected for Palomar~15 \citep{harris91}. The
concentration of 0.4 makes it less concentrated than any other
Galactic globular cluster. 

\begin{figure}
\includegraphics[width=0.48\textwidth]{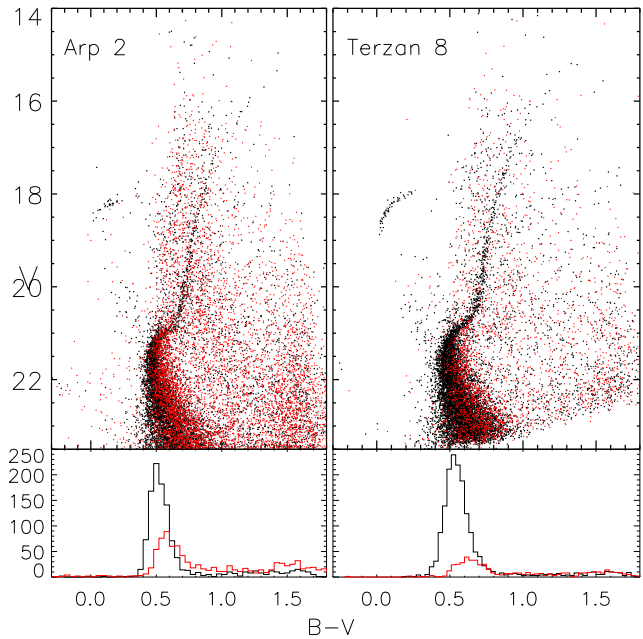}
\caption{CMD of stars inside (black) and outside (red) the measured
  tidal radius in Arp~2 and Terzan~8. The lower panels indicate
    the color distribution of both populations between V=22--22.5.}
\label{fig:extra}
\end{figure}

Color-magnitude diagrams of the extra-tidal regions of Arp~2 and
Terzan 8 can be seen in Figure~\ref{fig:extra} (red points),
overplotted on the inner regions (black dots). The presence of an
extra-tidal MS on both clusters is clearly revealed. Even though at
first sight this could mean an ill determination of the tidal radii,
the excellent fits of the King profile in Figure~\ref{fig:structure}
show no evidence of an enhancement of stars in the outer parts of the
clusters. A closer look to the extra-tidal MSs show that even though
the one in Arp~2 is slightly redder, but almost totally coincident
with the cluster's MS, the extra-tidal MS in Terzan 8 is clearly
redder. When we select stars between $V=$22 and 22.5
(Fig. \ref{fig:extra}, lower panels), the median color of the outer
components are 0.63 and 0.66 for Arp 2 and Terzan 8, respectively,
with a color dispersion of $\sim0.1$ in both cases. These median
values are 0.1 and 0.12 redder than the ``inner'' MS at the same
magnitud level. Even though a large amount of binaries in the
outskirts of the clusters or some strange pattern of differential
reddening affecting only the surroundings, but not the clusters
themselves, could serve as explanations, the most likely explanation is
the presence of the Sgr MS which is known to be more metal rich than
these clusters \citep[e.g.][]{layden00}, hence explaining its redder
MS. This Sgr MS should have a corresponding Sgr RGB which is not seen
in the CMDs. A likely explanation is the much lower spatial density of
the Sgr field; if we count the stars in the 22--22.5 magnitude range
(Fig. \ref{fig:extra}, lower panels), the density ratio between the
inner and outer populations is close to 5. Since the cluster's RGB are
difficult to notice without the coloring scheme in
Fig. \ref{fig:extra}, a possible Sgr RGB would be even harder to
detect. Wider and deeper studies of Arp~2 and Terzan~8 may reveal
their disruption into the Sgr body, although the separation with the
contamination from Sgr will be difficult, requiring a careful star
selection based on multi-color photometry.

\begin{figure}
\includegraphics[width=0.49\textwidth]{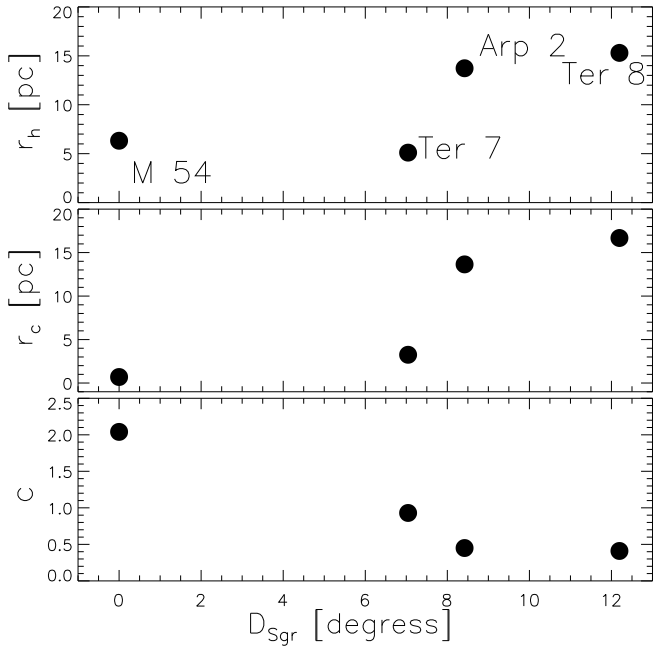}
\caption{Concentration\,--\,distance and core/half
  radius\,--\,distance relations for the bona fide Sgr clusters. M~54
  is placed at the center of Sgr (but see \citet{siegel11}). The data
  for M~54 and Terzan~7 come from \citet{harris96}, while those for
  Arp~2 and Terzan~8 from our measurements. }
\label{fig:dist_trends}
\end{figure}

\begin{table*}
 \centering
 \caption{Comparison of the structural parameters with literature values.}
\label{table:comparison}  
\begin{tabular}{@{}lllllll@{}}
  \hline
  Cluster &$r_{\rm c}$ (published)&References&$r_{\rm c}$ (this paper)&$r_{\rm t}$ (published) &References&$r_{\rm t}$ (this paper) \\
  &[arcmin]& &[arcmin]&[arcmin]& &[arcmin]\\
  \hline
  Arp~2     & 2.0, 1.67, 1.59, 1.64, 1.42        & 1, 3, 4, 7, 8 & $1.64\pm0.14$  &$\gtrsim5$, 16.7, 12.6, 10.7 & 1, 3, 4, 8 & $4.61\pm0.30$\\
  NGC~5634  & 0.24, 0.21, 0.09, 0.19             & 2, 3, 8, 9    & $0.70\pm0.07$  &8.35, 10.37                  & 3, 8       & $5.73\pm0.17$\\
  Palomar~12& 0.48, 0.53, 1.10, 0.03, 0.63, 0.02 & 1, 3, 4, 5, 6, 8 & $0.64\pm0.05$  &10.7, 10.5, 8.7, 2.5, 7.6, 15.28 & 1, 3, 4, 5, 6, 8 & $8.97\pm2.43$\\
  Terzan~8  & 1.0, 1.25                          & 4, 7          & $2.18\pm0.10$  &4.0                      & 4              & $5.56\pm0.19$\\ 
  \hline
\end{tabular}

\medskip
Reference list: 1 \citet{peterson76}; 2 \citet{kron84}; 3 \citet{chernoff89}; 4 \citet{trager93}; 5 \citet{trager95}; 6 \citet{rosenberg98}; 7 \citet{mackey03c}; 8~\citet{mclaughlin05}; 9 \citet{carballo11}
\end{table*}

We note that the half-light radii of the globular clusters lying in
the Sgr body show a clear increasing trend as a funcion of distance to
the Sgr center (Figure~\ref{fig:dist_trends}, top panel). A similar
relation has also been noticed in our Galactic globular cluster system
\citep[e.g.][]{vandenbergh00} and in many others
\citep[e.g.][]{vandenbergh00,barmby02,cantiello07,hwang11}. A similar
relation between central density and galactocentric distance has been
shown by \citet{penarrubia09}, although based on the observational
data from \citet{mackey03c}, which may not be appropriate for the more
extended clusters (see below). Core radii and concentrations also show
clear trends with respect to Sgr distance (Figure~\ref{fig:dist_trends},
middle and low panels). The inclusion of former Sgr clusters (like
NGC~5634 and Palomar~12) into these plots would increase the scatter
in the relations, although their original positions within Sgr are
unknown.

Core radius measurements for the four clusters can be found in the
literature with a rather large scatter, while tidal radius values are
scarcely available. Table \ref{table:comparison} compiles structural
parameters for the four clusters as found in the literature. In the
case of Arp~2, our core radius estimate is in good agreement with the
values previously measured, while our tidal radius is significantly
lower. This may be an effect of the strong contamination by the
Galactic disc in the Arp~2 field (see Figure~\ref{fig:cmds}); shallower
observations not reaching below the MS turnoff could be dominated by
this contamination, artificially increasing the tidal radius
determination.

In NGC~5634 all the core radii values are significantly smaller than
our measured $r_{\rm c} = 42\arcsec$. This is probably because our
photometry, and subsequently the star counts, are very incomplete in
the inner $\sim 30\arcsec$ of the cluster, implying an underestimate
of the central distribution. The extrapolation of the King profile
into these central parts probably underestimates the real central
light contribution, transforming into an overestimation the core
radius. 

Most of the core radii values of Palomar~12 are again smaller than our
determination. These very low values are reproduced in the
\citet{harris96} catalogue, spuriously setting Palomar~12 as the most
concentrated cluster with $c=2.98$. The explanation for our higher
$r_{\rm c}$ value comes perhaps from the sparse and poorly populated
nature of this cluster. As seen in the color-magnitude diagram in
Figure~\ref{fig:cmds}, the red giant branch (RGB) and sub giant branch
are poorly populated, so only deep observations reaching well below
the main sequence turn-off, as provided in this work, can provide
enough stars to measure structural properties confidently. Slightly
shallower and with a somewhat smaller field-of-view CCD observations
obtained by \citet{rosenberg98} give parameters in very good agreement
with ours, with $r_c=37.8$~pc and $c=1.08$. 

\begin{figure*}
\includegraphics[width=\textwidth]{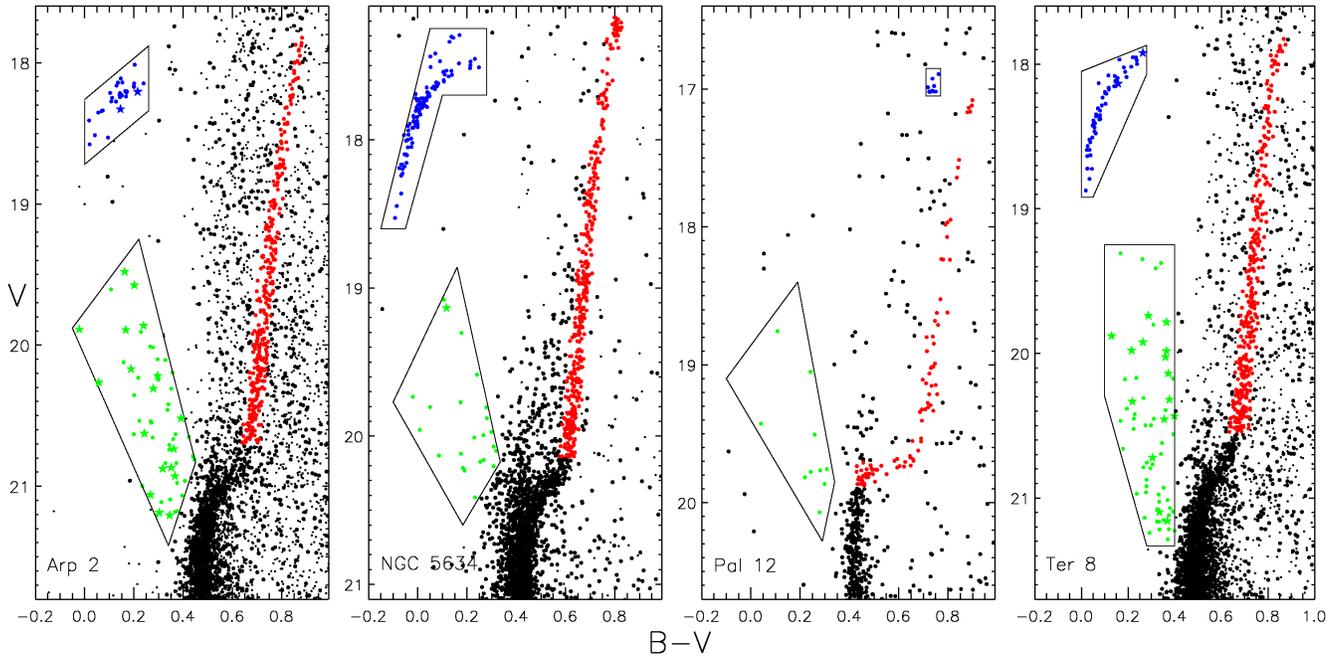}
\caption{Sub-populations selection criteria. BSs are marked
  with green symbols, RGB stars with red symbols and HB stars with
  blue symbols. The starry symbols in the BS and HB regions indicate stars
  outside the tidal radius of the cluster.}
\label{fig:boxes}
\end{figure*}

Finally, Terzan~8 has been scarcely studied. Values given by
\citet{trager93} are acknowledgedly a ``guess'', while
\citet{mackey03c}, based on an HST surface brightness profile
extending to 75\arcsec, give a core radius of 75.4\arcsec. Again, our
radially extended observations, although probably not as deep as the
HST imaging, give more confident results.

In summary, we believe our deep, homogeneous and radially extended
photometry gives a better base to fit King profiles on these clusters,
with the possible exception of NGC~5634, providing more accurate
measurements of their structural parameters.

\section{Blue stragglers distribution}
\label{sec:stragglers}

The spatial distribution of BSs in the Sgr clusters is studied
emphasizing its comparison with the distribution of RGB and HB
stars. The latter two types of stars are expected to have no
peculiarities in their radial distribution. The first step is to
determine which stars can be considered as part of the different
sub-populations in each cluster. The chosen color and magnitude
criteria are indicated in the boxes in Figure~\ref{fig:boxes}. We
found 55, 23, 10, and 62 BS candidates in Arp~2, NGC~5634, Palomar~12,
and Terzan~8, respectively, when considering the entire field of
view. The results presented below are mostly insensitive to slight
variations on the election of the red edge of the BS selection box,
especially in the most interesting cases of Arp~2 and Terzan~8.

\subsection{Estimating the contamination}

The relatively large studied area permits us to do an estimation of
the contamination introduced by field stars. We consider as ``field''
all the stars outside the tidal radius. Even though this criterion is
very conservative and some genuine cluster members could inhabit these
extra-tidal regions \citep[e.g.,][]{correnti11,walker11} and in that
case be rejected as contamination, the field of view does not allow us 
to use as a control field stars further away from the clusters. This
criterion is preferred over the star counts from Galactic models,
since the contamination from Sgr itself can be more properly taken into
account. Since the Sgr populations are known to vary within the
galaxy's body \citep[e.g.,][]{alard01,giuffrida10}, a local estimation
is also preferred over using a control field from another study.

\begin{table}
 \centering
 \caption{Sub-populations star counts. Numbers in parenthesis indicate the expected field contamination in each bin. Errors in the BS specific frequency have been calculated assuming Poissonian noise of uncorrelated variables \citep{ferraro06,beccari11}. The last column indicates the fraction of the cluster's luminosity in each bin.}
\label{table:bss}  
\begin{tabular}{@{}lllllc@{}}
  \hline
Range (\arcsec) &     $N_{\mathrm{BS}}$       &      $N_{\mathrm{HB}}$     &  $N_{\mathrm{RGB}}$  & $\frac{N_{\mathrm{BS}}}{N_{\mathrm{HB}}}$ &$\frac{L_{\mathrm{samp}}}{L_{\mathrm{tot}}}$\\
\hline
\multicolumn{6}{c}{\bf Arp~2}\\
\hline
0--64   &         10 (0.48)   &   8 (0.05) &  57 (3) &1.25 $\pm$ 0.44 &0.31\\
64--96  &         10 (0.60)   &   9 (0.07) &  48 (4) &1.11 $\pm$ 0.37 &0.18\\
96--198  &        10 (3.53)   &   10 (0.39)&  70 (22)&1.00 $\pm$ 0.32 &0.35\\
198--276  &       7  (4.36)   &    2 (0.48)&  35 (27)&3.50 $\pm$ 2.47 &0.16\\
\hline
\multicolumn{6}{c}{\bf NGC~5634}\\
%Range ($\arcsec$) &     BS      &       HB    &          RGB\\
\hline
30--40      &     8 (0.0)  &       28 (0.0)    &     40  (0.0)& 0.29 $\pm$ 0.05&0.10\\ 
40--114     &     8 (0.1)  &       61 (0.0)    &     230 (0.7)& 0.13 $\pm$ 0.01&0.41 \\
114--345    &     6 (1.0)  &       16 (0.0)    &     94  (6.7)& 0.38 $\pm$ 0.09&0.17\\
\hline
\multicolumn{6}{c}{\bf Palomar~12}\\
%Range ($\arcsec$) &     BS      &       HB    &          RGB\\
\hline
0--60      &      5 (0.03)   &      5 (0.0)   &       92 (0.23)& 1.0 $\pm$ 0.45&0.50\\
60--500    &      5 (2.22)   &      3 (0.0)   &       128 (15.5)& 1.7 $\pm$ 0.96&0.48\\
\hline
\multicolumn{6}{c}{\bf Terzan~8}\\
%Range (\arcsec) &      BS    &         HB     &         RGB\\
\hline
0--60       &     12 (0.45)  &      12 (0.3)  & 57 (1.7) & 1.0 $\pm$ 0.28 &0.14\\
60--110     &     12 (1.06)  &      12 (0.7)  & 74 (4.1) & 1.0 $\pm$ 0.28 &0.30\\
110--155    &     12 (1.47)  &      10 (1.0)  & 51 (5.8) & 1.2 $\pm$ 0.38 &0.23\\
155--340    &     12 (11.3)  &      11 (7.6)  & 94 (44.5)& 1.1 $\pm$ 0.33 & 0.33\\
\hline
\end{tabular}
\end{table}

Contamination is then estimated by counting stars outside the tidal
radius which satisfy the same color and magnitude limits shown in
Figure~\ref{fig:boxes} for each sub-population. The star counts are
then normalized to the areas defined by the radial ranges set in
Table~\ref{table:bss}. The sizes of these radial ranges have been
established in order to have approximately the same number of BSs
inside each of them. The estimated number of contaminating stars for
each sub-population in each radial bin is given in parenthesis in
Table \ref{table:bss}. The Galactic contamination in the magnitude and
color range of the BSs is expected to be low, so this contamination,
especially in the outer parts of the clusters Arp~2 and Terzan~8, must
come from Sgr itself; whether these stars are part of a young
population or BSs in the Sgr field remains an open issue
\citep[e.g.,][]{momany07}. In the case of Palomar 12, where the tidal
is larger than our field of view, the contamination has been assumed
to have the same spatial density calculated in the NGC 5634 field.

\subsection{The radial distribution of BSs}

\begin{figure}
\includegraphics[width=0.49\textwidth]{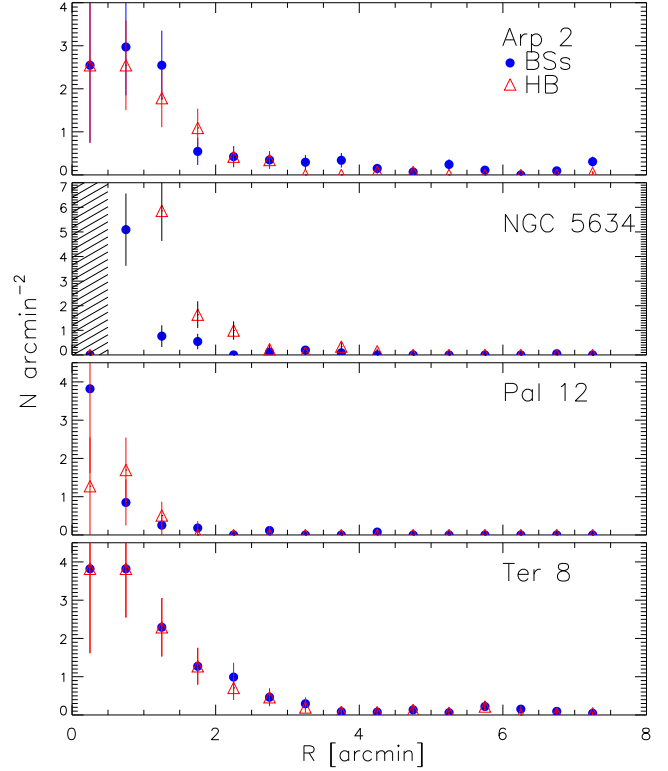}
\caption{Radial density distribution of BSs (filled blue circles) and
  HB stars (open red triangles) in the four clusters. The dashed lines
  indicate the radial range not included in the analysis of
  NGC~5634. Error bars are derived from an assumed Poissonian noise in
  the star counts. }
\label{fig:density}
\end{figure}

The radial density distribution of BSs in the four clusters can be
seen in Figure~\ref{fig:density}, compared to the distribution
of HB stars.

The BSs distribution in Arp~2 has already been studied by
\citet{buonanno95} and \citet{carraro11}.  Our study has as the main
advantage over these previous studies that the sampled area is $\sim$
24 and 10 times larger than the ones surveyed by \citet{buonanno95}
and \citet{carraro11}, respectively, allowing a study across the
entire cluster and a more proper determination of the field
contamination. A second difference is that we use a more conservative
red delimitation for the BSs zone in order to avoid interloping main
sequence stars. The BS radial distribution is not unlike the HB
distribution as can be seen in Figure~\ref{fig:density} (top panel),
when Poissonian errors are considered.

The BSs distribution in Terzan~8 was studied in the sample of
low-luminosity globular clusters of \citet{sandquist05}, but without
giving details for this specific cluster. The observed BS distribution
is completely indistinguishable from the HB distribution
(Figure~\ref{fig:density}, bottom panel).

In contrast to the smooth decline of the BS density profile in the
aforementioned clusters, NGC~5634 and Palomar~12 show a much more
peaky central distribution, with a sudden decline at $\sim\,1\arcmin$
(Figure~\ref{fig:density}, middle panels). A clear view of the innermost
($R<30\arcsec$) BS behaviour of NGC~5634 is not possible because of
the severe crowding.

\begin{figure}
\includegraphics[width=0.49\textwidth]{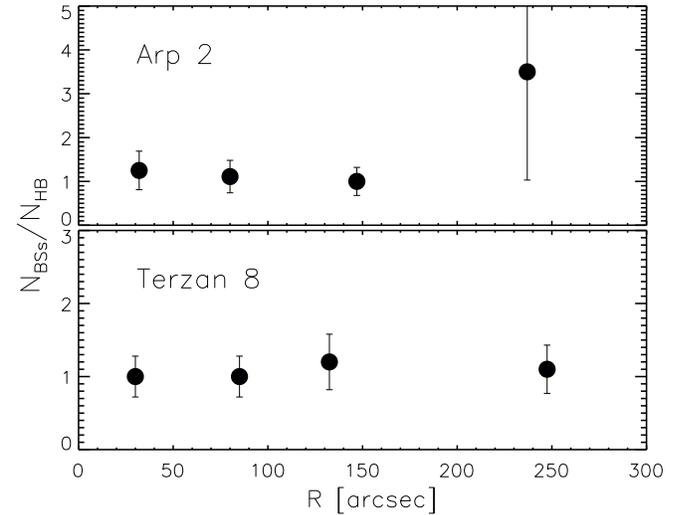}
\caption{BS specific frequency profiles in Arp 2 and Terzan 8.}
\label{fig:specific}
\end{figure}

\begin{figure}
\includegraphics[width=0.48\textwidth]{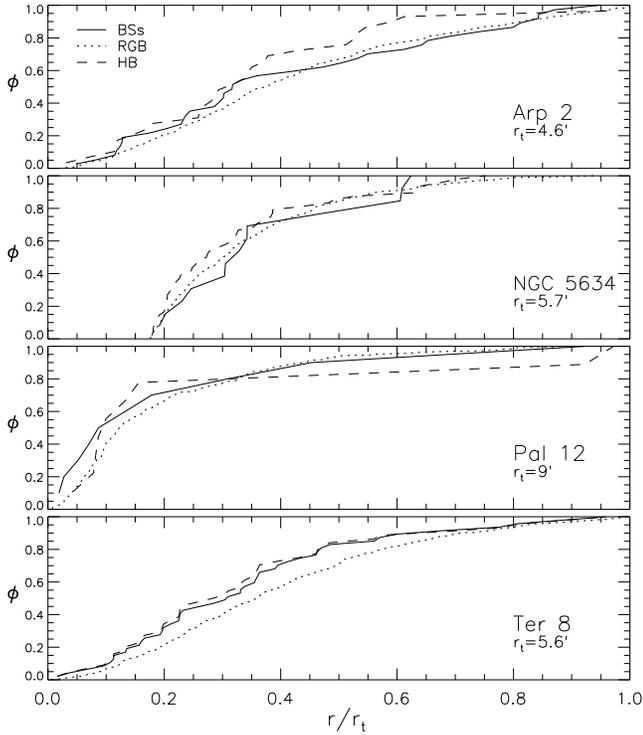}
\caption{Cumulative distributions of BSs (solid line), HB stars
  (dashed line) and RGB stars (dotted line) in the four clusters
  within each cluster's tidal radius.}
\label{fig:cumulative}
\end{figure}

A further indicator of segregation is the profile of the BSs specific
frequency, i.e. the number of BSs normalized to the number of HB
stars, $S_{\mathrm{BSs}}\equiv N_{\mathrm{BS}}/N_{\mathrm{HB}}$
(column 5 in Table~\ref{table:bss}). ``Normal'' clusters show a
bimodal radial behaviour, where the normalized BSs show a peak at the
center followed by a mid-radial zone of avoidance and an external
upturn \citep[e.g.][]{ferraro09b}. This U-shape is considered as a
signature of the sinking of the heavy BSs into the bottom of the
potential, producing the zone of avoidance and the central
concentration \citep[e.g.,][]{mapelli06}. NGC~5634 seems to follow the
expected U-shape trend, even though the rather low number of stars
precludes a firm conclusion. Palomar~12 shows a clear central density
of BSs, with a high value in the inner 30\arcsec (see
Figure~\ref{fig:density}), but again the low number of stars does not
allow us to see a bimodal distribution.  Arp~2 shows a mostly flat
$S_{\mathrm{BS}}$ profile, although a hint of a central concentration
can be seen (Fig. \ref{fig:specific}, top panel). Most impressive is
the external upturn of $S_{\mathrm{BS}}=3.5\pm2.5$. If we consider the
lowest possible value, 1.0, then $S_{\rm BS}$ shows no external upturn
and the profile is mostly flat. The value of $S_{\mathrm{BS}}=3.5$, on
the other hand, is higher than the one seen for almost any globular
cluster \citep[with the exception of some very faint clusters like E3
and Palomar~13, studied by][]{sandquist05}, but in good agreement to
the value obtained for the Galactic halo \citep{preston00}. If the
contamination is mainly produced by Sgr BSs instead of a young
population as postulated by \citet{momany07}, it would indicate that
the Sgr BSs specific frequency is the same as the one of the Galactic
halo, although this goes against the specific frequency
$S_{\mathrm{BS}}=0.55$ for the Sgr field calculated by
\citet{momany07}. If we consider the Arp~2 ``field'', i.e. the stars
beyond the cluster's tidal radius, the specific frequency jumps to an
unprecedented value of 9, which can be considered as an indication
against a Sgr BS population. A rather high value is also seen in the
outer parts of Palomar~12 (see Table~\ref{table:bss}), although again
small-number statistics precludes us from assigning it any
relevance. Finally, Terzan~8 shows a totally flat $S_{\mathrm{BS}}$
profile (Fig. \ref{fig:specific}, bottom panel). The Terzan~8
``field'' has $S_{\mathrm{BS}}=1.6$, with 8 HB stars lying beyond the
tidal radius, perhaps a further indication of possible disruption.

To test more quantitatively whether the radial distribution of BS, HB
and RGB stars are extracted from the same parent distribution,
indicating an absence of segregation, we use the k-sample
Anderson-Darling test \citep[][hereafter A-D test]{scholz87} as
implemented in the R programming language \citep{scholz11}. The A-D
test is similar to the more widespread Kolmogorov-Smirnov test, but
with greater sensitivity to the tails of the cumulative distribution
(Figure~\ref{fig:cumulative}). In the case of Arp~2, the A-D test
indicates that the probability that BSs, HB and RGB stars come from
the same distribution is less than 1\%. This is not totally surprising
if we consider the contamination as a function of radius. Star counts
in Table~\ref{table:bss} indicate that from the 7 BSs found beyond
$\sim$ 200\arcsec around 4 could come from field contamination. The
A-D probability changes dramatically when this last radial bin is
removed from the analysis. If only stars with $R\lesssim200\arcsec$
are considered, the probability that BS, HB and RGB stars come from
the same distribution rises to 66\%, i.e. no significant evidence for
segregation. As a comparison, the A-D probability that HB and RGB
stars are extracted from the same distribution is 58\% in the same
radial range.

Although less affected by contamination than Arp~2 (see
Figure~\ref{fig:cmds}), the last radial bin in Terzan~8 (see
Table~\ref{table:bss}) indicates that it is even possible that all BSs
in this range are from the field instead of the cluster. Considering
only stars inside $R \lesssim 160\arcsec$, the A-D test gives a
probability of 75\% that BS, HB and RGB stars come from the same
distribution, strongly disfavouring any mass segregation as in
Arp~2. This number can be compared to the 56\% probability that HB and
RGB stars come from the same distribution. For the clusters outside
the Sgr body, the probabilities are lower. While for Palomar~12 the
A-D test gives a probability of 44\% that the sub-samples have the
same radial distribution, this number drops to less than $10^{-4}$\%
for NGC~5634, strongly favoring the existence of mass segregation. The
low number of BSs in Palomar~12 prevents us from considering this
number as clear-cut evidence for an absence of mass segregation~-- the
same result obtained by \citet{rosenberg98}.

Even though the effect of field contamination is partially avoided in
Arp~2 and Terzan~8 by excluding the outer radial bin from the
analysis, contamination is not completely negligible in the inner
bins. To test how robust the results from the A-D test are against
sample contamination in these two clusters, we use the estimations
provided in Table~\ref{table:bss}: from the stars lying in each radial
range we randomly subtracted the expected contamination, generating
100 new samples for the radial distribution of BS, HB and RGB stars
in each cluster. Each triad of distributions was tested with the A-D
test. The mean value of the probability that the three samples in Arp~2
come from the same distribution slightly decreases to 60\%, from the
original 67\%. In the case of Terzan~8, the same procedure indicates
that the probability is 74\%, a negligible decrease from the original
75\%. These estimates imply that the absence of segregation indicated by
the A-D test in the inner $\sim$ 3 \arcmin of Arp~2 and Terzan~8 is a 
result that is not affected by field contamination.

\section{Discussion}
\label{sec:discussion}

Our analysis of the BS distribution shows that Terzan~8,
and probably Arp~2, have not relaxed yet. This is not altogether
surprising, since relaxation time depends on the size of the dynamical
system and, as seen in Sect.~\ref{sec:structure}, these two clusters
have large cores. We calculate the central relaxation times of the
four clusters using Eq. 10 from \citet{djorgovski93}, assuming a mean
stellar mass of $1/3\,M_{\odot}$, adopting cluster distances from
\citet{harris96} and taking the core radii values calculated in
Sect.~\ref{sec:structure}. Results can be seen in the last column of
Table~\ref{table:king}. Terzan~8 has a central relaxation time of
$t_\mathrm{crt}=16.6$~Gyr~-- significantly larger than the cluster's
age. Arp~2 has $t_\mathrm{crt}=7.9$~Gyr. This is a factor $\sim1.7$
shorter than the age of the oldest clusters in the Galaxy, and
indicates that if Arp~2 were as old as them, mass segregation should
already be visible. Since according to the A-D test the BS distribution 
and the specific frequency profile show at most weak evidence of
segregation, this could imply that Arp~2 is indeed a slightly young
($\sim 9$~Gyr) cluster, as first postulated by \citet{buonanno95}.

Arp~2 and Terzan~8 are among the globular clusters with the lowest
concentration measured. Since in these environments direct collisions
and binary disruption are expected to be negligible, the BS
properties should be more similar to the field than to more
massive and concentrated clusters. Nevertheless, the specific frequency
of BSs in the local halo is found to be significantly higher, with
$S_{\mathrm{BS}}\sim 4$ \citep{preston00}, much higher than in the
lowest-concentration clusters, suggesting that the specific frequency
is primarily driven by the total luminosity of the cluster \citep{sandquist05},
instead of concentration. Notably, the BSs specific frequency in
these clusters is within the range estimated for the Galactic bulge,
namely $0.31<S_{\mathrm{BS}}<1.23$ \citep{clarkson11}.

\section{Summary}
\label{sec:summary}

We have presented $BV$ photometry of four globular clusters associated
with the Sagittarius dwarf: Arp~2, NGC~5634, Palomar~12 and Terzan~8.
Arp~2 and Terzan~8 have remarkably low concentrations and with
half-light radii of $\sim$15 pc, not unlike the ``extended star
clusters'' now commonly found in Local Group galaxies \citep[e.g.][and
references therein]{hwang11}. Terzan~8 presents clear evidence of a
non-segregated BS population, indicating its dynamical youth and
joining in this condition much more massive clusters as $\omega$~Cen
\citep{ferraro06} and NGC~2419 \citep{dalessandro08b}, along with the
seemingly dissolving Palomar~14 \citep{beccari11,sollima11}. Although
not as strongly as Terzan~8, Arp 2 also shows at best weak evidence of
mass segregation, supporting the idea that it is a relatively young
globular cluster, in spite of its predominantly blue HB morphology.

\section*{Acknowledgments}
We thank the referee, Antonio Aparicio, for a helpful report which
improved the clarity of our results.  We thank Juan V\'eliz at the
P. Universidad Cat\'olica de Chile for help with the data handling. We
thank Giacomo Beccari for useful suggestions and Karla
\'Alamo-Mart\'inez for her advice on the usage of the R programming
language. LJ acknowledges the support by the 2-year ESO PhD
studentship, held in ESO, Santiago, as well as by grant
No. 205/08/H005 (Czech Science Foundation) and MUNI/A/0968/2009
(Masaryk University in Brno). This project is supported by the Chilean
Ministry for the Economy, Development and Tourism's Programa
Iniciativa Cient\'{i}fica Milenio through grant P07-021-F, awarded to
The Milky Way Millennium Nucleus; by the BASAL Center for Astrophysics
and Associated Technologies (PFB-06), by the FONDAP Center for
Astrophysics (15010003), by Proyecto Fondecyt Regular \#1110326, and
by Proyecto Anillo ACT-86. This research has made use of the SIMBAD
database, operated at CDS, Strasbourg, France.

\bibliographystyle{mn2e}
\bibliography{sgr}

\bsp

\label{lastpage}

\end{document}